\documentclass{aa}
\usepackage{color}
\usepackage{graphicx}
\usepackage{txfonts}
\usepackage[colorlinks, citecolor=blue]{hyperref}
\usepackage{mathtools}
\usepackage{ulem}
\usepackage{natbib}
\bibpunct{(}{)}{;}{a}{}{,} 

\newcommand{\GRALGL}{GRAL113100-441959}
\newcommand{\zs}{z_{\text{s}}} % redshift of the source
\newcommand{\zl}{z_{\text{l}}} % redshift of the lens
\newcommand{\xgal}{\theta_{\text{gal},1}} % lens position x
\newcommand{\ygal}{\theta_{\text{gal},2}} % lens position y
\newcommand{\bx}{\beta_{1}} % src position x
\newcommand{\by}{\beta_{2}} % src position y
\newcommand{\tE}{\theta_{\text{E}}} % Einstein radius
\newcommand{\tmax}{\sub{\theta}{max}}
\newcommand{\tq}{\theta_{q}} % q position angle
\newcommand{\tg}{\theta_{\gamma}} % shear position angle
\newcommand{\Ndof}{\sub{N}{dof}}

\newcommand{\bb}{\boldsymbol{\beta}}
\newcommand{\bt}{\boldsymbol{\theta}}
\newcommand{\bp}{\boldsymbol{p}}

\newcommand{\sub}[2]{#1_{\text{\scalebox{.9}{#2}}}}
\newcommand{\subb}[3]{#1_{\text{\scalebox{.9}{#2}},#3}}
\newcommand{\arc}[2]{#1\!\!^{\prime\prime}#2}
\newcommand{\dek}[2]{#1\!\!^{\circ}#2}

\newcommand{\ba}{\boldsymbol{\alpha}}
\newcommand{\bha}{\hat{\ba}}
\newcommand{\bta}{\tilde{\ba}}

\newcommand{\kp}{\kappa}
\newcommand{\hkp}{\hat{\kp}}

\newcommand{\bhb}{\hat{\bb}}

\begin{document}

%   \title{Gaia GraL: {\it Gaia} DR2 Gravitational Lens Systems. IV. \\GRAL113100-441959 confirmation and system modelling}
   \title{Gaia GraL: {\it Gaia} DR2 Gravitational Lens Systems. IV. \\Keck/LRIS spectroscopic confirmation of GRAL113100-441959\\ and model prediction of time-delays 
   %\textcolor{red}{[CD: What about : A new quadruply imaged quasar : GRAL113100-441959, spectroscopic confirmation and modelisation.]} 
   }
   \author{O. Wertz\inst{1},
   		   D. Stern\inst{2},
   		   A. Krone-Martins\inst{3},
   		   L. Delchambre\inst{4},
           C. Ducourant\inst{5},
           F. Mignard\inst{6},
           R. Teixeira\inst{7},\\
           L. Galluccio\inst{6},
           J. Kl\"{u}ter\inst{8},
           S.G. Djorgovski\inst{9},
           M.J. Graham\inst{9},
           J. Surdej\inst{4},
		   U. Bastian\inst{8},\\
           J. Wambsganss\inst{8,10},
           C. Boehm\inst{11},
           J.-F. LeCampion\inst{5},
           E. Slezak\inst{6}
        }

  \institute{
         Argelander-Institut f\"{ u}r Astronomie, Universit\"{ a}t Bonn,  Auf dem H\"{ u}gel 71, 53121 Bonn, Germany\\
          \email{owertz@astro.uni-bonn.de}
          \and
        Jet Propulsion Laboratory, California Institute of Technology, 4800 Oak Grove Drive, Pasadena, CA 91109, USA
          \and
  		 CENTRA, Faculdade de Ci\^encias, Universidade de Lisboa, Ed. C8, Campo Grande, 1749-016 Lisboa, Portugal
          \and
         Institut d'Astrophysique et de G\'{e}ophysique, Universit\'{e} de Li\`{e}ge, 19c, All\'{e}e du 6 Ao\^{u}t, B-4000 Li\`{e}ge, Belgium
         \and
          Laboratoire d'Astrophysique de Bordeaux, Univ. Bordeaux, CNRS, B18N, all{\'e}e Geoffroy Saint-Hilaire, 33615 Pessac, France
          \and
         Universit\'{e} C\^{o}te d'Azur, Observatoire de la C\^{o}te d'Azur, CNRS, Laboratoire Lagrange, Boulevard de l'Observatoire, CS 34229, 06304 Nice, France
         \and
         Instituto de Astronomia, Geof\'isica e Ci\^encias Atmosf\'ericas, Universidade de S\~{a}o Paulo, Rua do Mat\~{a}o, 1226, Cidade Universit\'aria, 05508-900 S\~{a}o Paulo, SP, Brazil
         \and
        Zentrum f\"{u}r Astronomie der Universit\"{a}t Heidelberg, Astronomisches Rechen-Institut, M\"{o}nchhofstr. 12-14, 69120 Heidelberg, \\Germany
        \and
         California Institute of Technology, 1200 E. California Blvd, Pasadena, CA 91125, USA
         \and
        International Space Science Institute (ISSI), Hallerstra\ss e 6, 3012 Bern, Switzerland
        \and
         Department of Physics, University of Sydney, Australia
          }

   \date{Received October 6, 2018; accepted ???, ???}

%%%%% ABSTRACT %%%%%
  \abstract
   {%\textcolor{red}{[CD: you made the choice to cancel the structure of abstract, why ? I would not start the abstract by the description of the GraL group and the way the target was discovered but would start immediately with the object of interest.]} 
% context heading (optional)
   %\textcolor{green}{[OW: in addition to be incomplete, the abstract requires a serious revision]} 
   We report the spectroscopic confirmation and modeling of the quadruply imaged quasar \GRALGL, the first gravitational lens (GL) to be discovered mainly from astrometric considerations.
   %We report the spectroscopic confirmation of the gravitational lens \GRALGL, 
   %Previously discovered by the {\it Gaia GraL} group from cross-matching a merged list of quasars and robust quasar candidates with the {\it Gaia} Data Release 2 and the adoption of supervised machine learning methods, 
    Follow-up spectra obtained with 
   Keck/LRIS reveal the lensing nature of this quadruply-imaged quasar with redshift $\zs = 1.090 \pm 0.002$, but show no evidence 
   of the central lens galaxy. 
   Using the image positions and $G$-band flux ratios provided by {\it Gaia} Data Release 2 as constraints, we model the system with 
   a singular power-law elliptical mass distribution (SPEMD) plus external shear, to different levels of complexity. We show that relaxing the 
   isothermal constraint of the SPEMD is not statistically significant, and thus we simplify the SPEMD to a singular isothermal ellipsoid to 
   estimate the Einstein radius of the main lens galaxy $\tE = 0.\!\!^{\prime\prime}851$, the intensity and position angle of the external 
   shear $(\gamma,\theta_{\gamma}) = (0.044,11.\!\!^{\circ}5)$, and we predict the lensing galaxy position to be  
   $(\xgal,\ygal) = (-0.\!\!^{\prime\prime}424,-0.\!\!^{\prime\prime}744)$ with respect to image A. We provide time 
   delay predictions for pairs of images, assuming a plausible range of lens redshift values $\zl$ between $0.5$ and $0.9$. We finally examine the impact 
   on time delays of the so-called Source Position Transformation, a family of degeneracies existing between different lens density profiles 
   that reproduce most of the lensing observables equally well. We show that this effect contributes significantly to the time delay error budget and cannot be ignored during the modelling. 
   This has implications for robust cosmography applications of lensed systems. \GRALGL\ is the first in a series of seven new spectroscopically confirmed GLs discovered from {\it Gaia} Data Release 2.
   }
   
  % aims heading (mandatory)
   %{}
  % methods heading (mandatory)
   %{}
  % results heading (mandatory)
   %{}
   %{} 
   
% {\it Gaia} Data Release 2 ({\it Gaia} DR2)
%, and compared their ability to reproduce the image positions and $G$-band flux ratios, both available in the {\it Gaia} Data Release 2. 
% does not improve substantially the goodness of the fit, while reducing the degrees of freedom by one. 


\titlerunning{{\it Gaia} GraL IV -- Gaia DR2 Gravitational Lenses}
\authorrunning{O. Wertz et al.}
\maketitle

%\textcolor{red}{[OW: In the case we include 3 more confirmed GL, the end of the introduction and the structure of the paper should be reconsidered. Spectra --> one image composed of several panels, each of them dedicated to one GL and showing the spectra of the images. The same idea for direct 2D image of the GLs (from DK1.54 of other). The `Lens confirmation` section should be divided in subsection, each of them dedicated to one GL, including details like source (and lens) redshift(s), configuration (double, quad), typical image separations, ... The latter can be quite short.]}

%%%%% INTRODUCTION %%%%%
\section{Introduction}

\begin{figure*}[!htp]
\begin{center}
\includegraphics[clip, trim=0.cm 6.5cm 0.cm 4cm, width=1.0\textwidth]{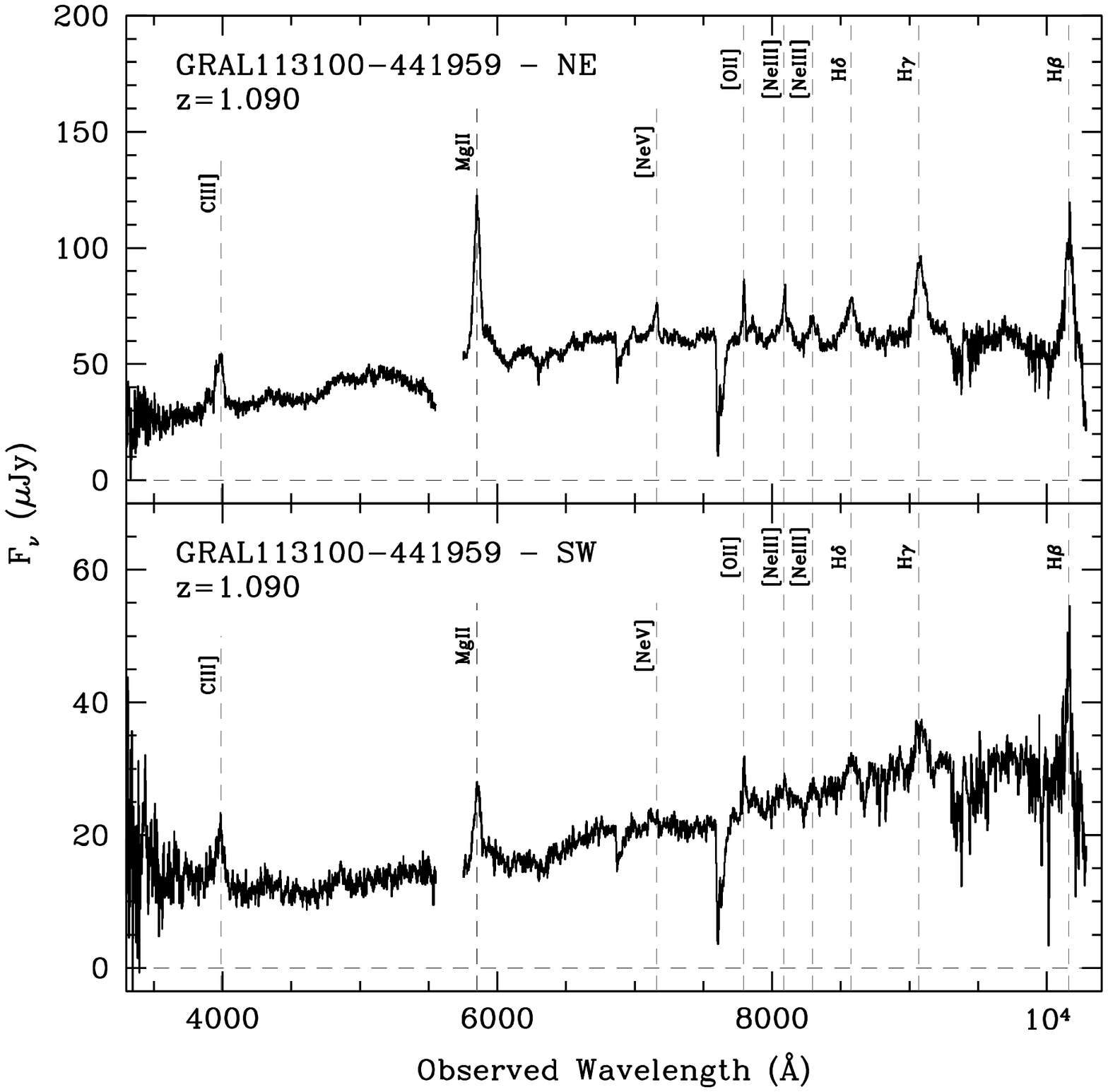}
\caption{Keck/LRIS spectra of images (A+B) and (C+D) at position angle $60^{\circ}$. Dashed lines identify emission lines used to confirm the lensing nature of \GRALGL.  %\textcolor{red}{[CD: naive question : is there any signature of the presence of two sources in the spectra, then we should mention or is it undetectable then we should maybie point out ?. AKM: No, as the two sources at each spectra have also the same spectra, they are indistinguishable.]} 
\label{fig:spectra}}
\end{center}
\end{figure*}

\begin{figure}[!htp]
\begin{center}
\includegraphics[width=0.49\textwidth]{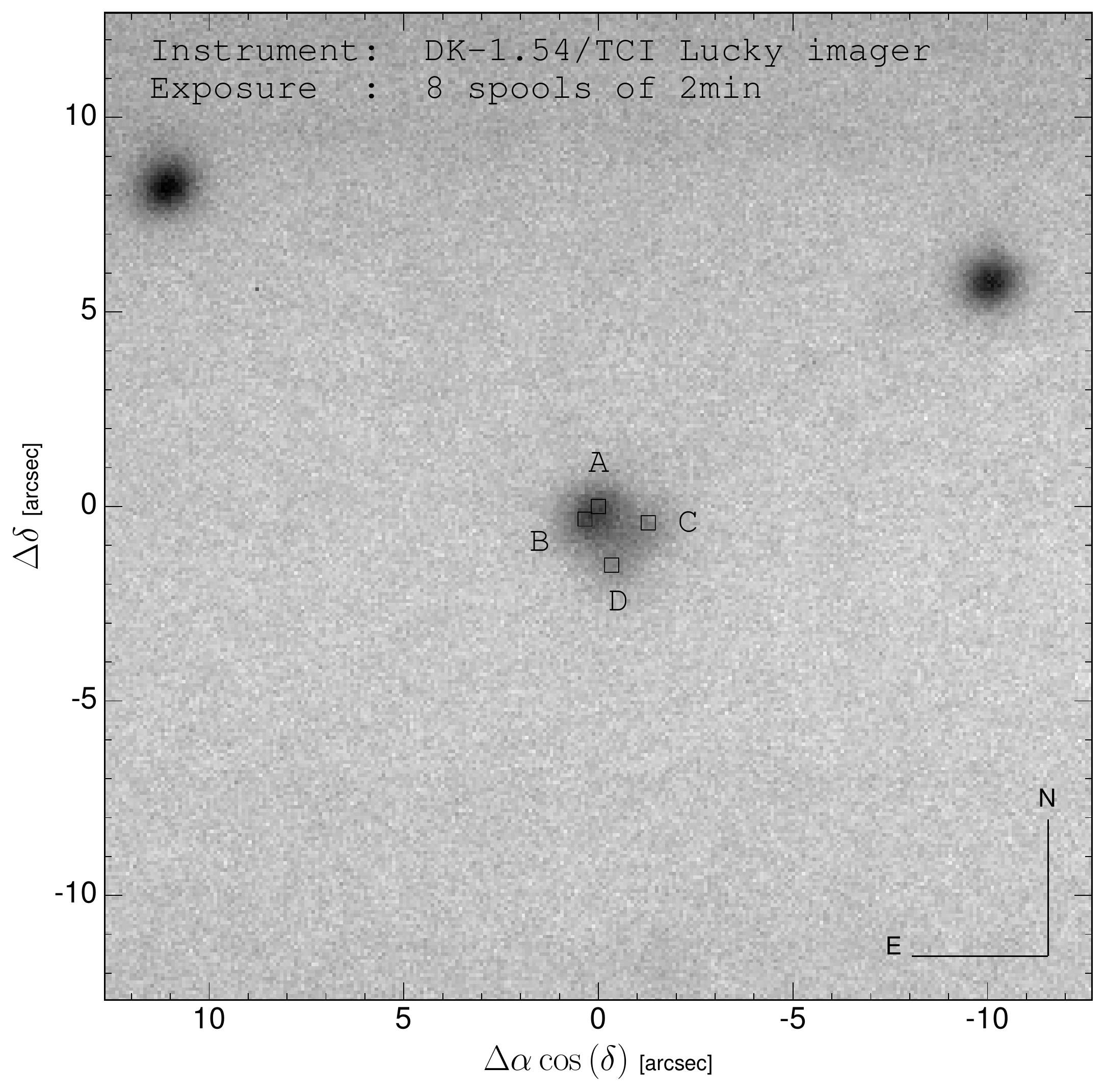}
\caption{The first direct imaging of \GRALGL\ obtained with the DK-$1.54$/TCI Lucky imager during the night of UT 2018 July 31. The black squares locate the lensed imaged positions as reported in the {\it Gaia} DR2.
\label{fig:DK154}}
\end{center}
\end{figure}

% >>> Notes and comments 
%\noindent
%\textcolor{red}{-- Provide a short text to explain why it is important to discover new GL.}

%\noindent
%\textcolor{red}{-- Presentation of our group, short summary of the methodology we applied to discover the candidates published in paper I.}

%\noindent
%\textcolor{red}{-- Explain why this candidate is a strong one, hence why we have decided to observe this one in particular.} 

%\noindent
%\textcolor{green}{[OW: Currently, the introduction is for the most part inspired by the MUSE proposal I've written recently.]} 
% e.g. the Gaia's observables related to this candidate are compatible with the the ones we have about already confirmed and well-known GLs

%\noindent
%\textcolor{green}{[OW: Populate the introduction with missing references --> Agnello and Co.]}

% >>> Main text
%\textcolor{red}{[CD: General remak : we should be coherent on the naming of objects. I already discussed with LD about the number of significant digits included in the names of paper III and here the naming between paper I and III are not agreeing for the present object. The names given in paper I indicate positions with 15$\arcsec$ precision in RA, this is not enough on my opinion. Do you have ideas about that ? TBD in next Liege meeting.]}
%Strong gravitational lenses (GLs) are among the most remarkable, exciting and useful observable extragalactic phenomena. 
Already suspected before General Relativity \citep{Einstein_GR_1916}, it was only after Einstein's final formulation of its theory that strong gravitational lensing (GL) was described quantitatively.
Then it took no less than three-quarters of a century to obtain a definitive observational proof when \cite{WCW_firstGL_1979} discovered a pair of quasars separated by $6$ arcseconds, with identical colors, redshifts, and spectra, thereby confirming the first doubly-imaged quasar.
Because the study of GLs constitutes a unique tool in various fields of astronomy \citep[see, e.g., ][and references therein]{Treu_TDcosmography_2016, Gilman2017, Jauzac2018, Zavala2018, Tagore2018}, they are highly sought, but not without difficulty. Even in this era of all-sky surveys, their discovery remains a great challenge, with barely a few hundred systems currently confirmed. A state-of-the-art list of currently known GLs can be found in \citet[][hereafter paper II]{Ducourant_GaiaGraL_paper_II_2018}. 

Data from the ESA/{\it Gaia} space mission \citep{2016A&A...595A...1G} is expected to change the situation dramatically. {\it Gaia} is conducting the largest, most precise, most accurate all-sky astrometric survey from space. Its main goal is to chart a three-dimensional map of our Galaxy based on measurements of parallaxes, proper motions, positions and spectro-photometric parameters for more than a billion stars. With an order of magnitude improvement over typical {\it HST} astrometric accuracy, {\it Gaia} will also detect $\sim600$,$000$ quasars \citep{Mignard_QSO_Gaia_2012, 2012A&A...543A.100R}, of which $2900$ are expected to be multiply-imaged and resolved in the final {\it Gaia} Data Release ({\it Gaia} DR), including $250$ systems with more than two lensed images \citep{2016Finet}.

As part of a larger effort to discover and study multiply-imaged quasar systems hidden in the heart of {\it Gaia} DRs, the {\it Gaia GraL} group has recently developed and successfully applied various techniques to identify new highly probable gravitational lens candidates from {\it Gaia}'s data. Our strategy is twofold and can be summarized as follows. Initially, our research focused on all known quasars that we compiled in a state-of-the-art list populated primarily with the Million Quasars Catalog \citep{2015PASA...32...10F, Flesch_MILLIQUAS_2017}, searching for the presence in {\it Gaia} DR2 of one or more nearby $(< 6\arcsec)$ point-like companion(s). This was the initial approach we took in \citet[][]{2018A&A...616L..11K}, hereafter paper I. %\textcolor{magenta}{and then applying supervised machine-learning techniques, where we discovered three quadruply-imaged lensed QSO candidates} \citep[][hereafter paper I]{2018A&A...616L..11K}. 
Next, we designed a dedicated method %called \texttt{SELenA} (Systematic Exploration of Lenses using Astrometry)  
%\textcolor{red}{[CD: As far as I know We never published anything with this SELENA denomination which has been now superseeded by GraL, I would drop this SELENA acronym. AKM: We can drop, but SELENA is a method, GraL is a group, so it was not superseeded. :-) ]} 
to blindly identify clusters of point-like objects from the information available in the {\it Gaia} DRs using the Hierarchical Triangular Mesh technique \citep[][]{Kunszt_HTM_2001}. This was the approach we took in \citet[][]{Delchambre_GaiaGraL_paper_III_2018}, hereafter paper III.

%\citet[][hereafter paper III]{Delchambre_GaiaGraL_paper_III_2018}
%and also supervised machine learning algorithms. 
%

The list of clusters generated from these two approaches is expected to be polluted with contaminants, resulting primarily from chance alignements of unrelated sources. To discard the most obvious ones, we thus applied soft astrometric filters to differentiate genuine candidates from fortuitous clusters of stars by studying the image proper motions and parallaxes of known lenses, as measured by {\it Gaia} (see paper II). 
{\it Gaia} DR2 also provides broad band photometric measurements in the $G$-band ($330$ to $1050$ nm), in particular, a color indicator derived from the integrated flux of the low-resolution blue photometer (BP, $300 - 680$ nm) and red photometer (RP, $630 - 1050$ nm) spectra \citep{Jordi_Gaia_2010, Evans_DR2_2018}. Because the gravitational lensing phenomenon is achromatic, we also rejected clusters for which the individual component BP$-$RP color indicators significantly differ from each other. 
With the purpose of considering only the most plausible candidates, we classified the remaining clusters that successfully passed the astrometric and photometric filters with respect to their chance of being a multiply-imaged quasar candidate. To this end, we assigned to each of them a probability that reflects the match between a candidate and the learning set composed of more than $10^8$ simulated image configurations that we used to build Extremely Randomized Trees \citep{Geurts_ERT_2006}. When considering a non-singular isothermal ellipsoid (NIE) plus external shear as the lens model, various tests have shown this method to be efficient in identifying known GLs from fortuitous clusters of stars with a detection probability of $97\%$ in the case of configurations with four lensed images along with a contamination ratio of $1.37\%$. %\textcolor{red}{[LD: 97\% with a contamination of 1.37\% or 90\% with a contamination of 0.46\%, as you prefer]}. 
By implementing this strategy to {\it Gaia} DR2, we discovered $15$ new highly probable quadruply-imaged quasar candidates, recently presented in paper III.
%%\citet[][hereafter paper III]{Delchambre_GaiaGraL_paper_III_2018}
Furthermore in this blind search, we also found $17$ well-known quadruply-imaged quasars for which three or four components are detected in {\it Gaia} DR2. This constitutes additional evidence of the robustness of our methodology. %tends to confirm 

\GRALGL\ was identified for the first time as a new highly probable GL candidate in paper I, % (second chart in Fig. 1), 
and was then rediscovered independently with an ERT probability of $96\%$ from the blind search technique presented in paper III 
(candidate number $[12]$ their Fig. 3). This is likely the first gravitationally lensed quasar discovered from astrometric/photometric survey data
. Prior to its spectroscopic confirmation, we were not able to perform any visual inspection of this candidate because the available surveys either lack spatial resolution and/or sensitivity (e.g., SkyMapper, \citet{Wolf_Skymapper_DR1_2018}, and ALLWISE, \citet{Wright_WISE_2010}), or they lack spatial coverage \citep[e.g., Pan-STARRS,][]{Chambers2016}. The limited spatial resolution of current all-sky southern surveys probably explains why \GRALGL\ remained unnoticed so far, confirming that a large number of GLs that can be observed from the ground still waits to be discovered.

In Sect. \ref{sec:obsdata} we describe the spectroscopic observations and confirm the lensing nature of \GRALGL. In Sect.\ref{sec:modeling}, we describe
the details of the simple lens modeling, and provide predictions for the time delays between pairs of lensed images in Sect.\ref{sec:tds}. Finally, we summarize our findings and conclude in Sect.\ref{sec:conclusions}.
%explore the impact of some lensing degeneracies on the model predicted time delays between pairs of lensed images. 

%%%%% LENS CONFIRMATION %%%%%
\section{Lens confirmation}
\label{sec:obsdata}

\subsection{Observations}

\noindent
On UT 2018 May 13, we observed \GRALGL\ with the dual-beam Low
Resolution Imaging Spectrometer \citep[][]{Oke_LRIS_1995} %(LRIS; Oke et al. 1995, PASP, 107,375) 
on the Keck~I telescope.  The conditions were photometric, and
we obtained two $300$s spectra, at position angles (PAs) of $60^{\circ}$
and $135^{\circ}$.  We used the 1\arcsec\ width slit, the 5600\AA\
dichroic, the 600~$\ell$~mm$^{-1}$ blue grism ($\lambda_{\rm blaze}
= 4000$~\AA), and the 400~$\ell$~mm$^{-1}$ red grating ($\lambda_{\rm
blaze} = 8500$~\AA).  This instrument configuration covers the full
optical window at moderate resolving power, $R \equiv \lambda /
\Delta \lambda \approx 1100$.  The observations were processed
using standard techniques within \texttt{IRAF}, and flux-calibrated using
observations of the spectrophotometric white dwarf standard stars
Feige~34, Feige~67, and Wolf~1346 obtained on the same night.

The target appears as a single source in the PA $=135^{\circ}$ observation, which was aligned along the brighter NE components of the lens.  Though the two components are not differentiated in the spectroscopy, the source is clearly spatially resolved, with a FWHM of $\sim \arc{1.}{4}$ compared to the $\sim \arc{1.}{0}$ seeing.  

In the PA $=60^{\circ}$ observation, the target is clearly resolved into two sources separated by $\sim 1\farcs2$ with identical spectroscopic features. One component is significantly brighter than the other.  Fig.\ref{fig:spectra} presents the spectra of the brighter (NE) and fainter (SW) components, each extracted with $\arc{0.}{5}$ box width. 
Based on Gaussian fits to the typical broad, quasar emission lines such as \ion{C}{III}]~$\lambda 1909$, \ion{Mg}{II}~$\lambda 2800$, and Balmer transitions of hydrogen, we measured a redshift of $z=1.090$ with a conservative estimate of the uncertainty of $0.002$.
%Both spectra clearly show a $z = 1.090 \pm 0.002$ quasar with typical broad, high-ionization emission lines such as \ion{C}{III}]~$\lambda 1909$, \ion{Mg}{II}~$\lambda 2800$, and Balmer transitions of hydrogen.

On UT 2018 July 31, \GRALGL\ was also observed with the $\arc{0.}{09}/\text{px}$ Two Color Instrument (TCI) Lucky imager \citep[][]{Evans_DK154_2016} mounted on the $1.54$m Danish telescope at La Silla, Chile. A sequence of eight spools of two minute exposures was obtained with the \texttt{RED} color channel. We combined all the quality bins of each spool to generate the $16$ min total exposure image shown in Fig.\ref{fig:DK154}.
%observations of \GRALGL with the 1.54m Danish telescop
%we also obtained the first direct imaging of \GRALGL\ 
%

\subsection{The lensing nature of \GRALGL}
\label{subsec:lensing_nature}

\noindent
Fig.\ref{fig:spectra} presents the calibrated spectra of \GRALGL\ from the PA
$=60^{\circ}$ observation; a wide aperture was extracted containing
both observed components.  The PA $=135^{\circ}$ observation is identical to the NE component.
The source is clearly identified as a quasar at $z_s = 1.090$ based on
strong detections of broad emission from \ion{C}{III}]~$\lambda 1909$,
\ion{Mg}{II}~$\lambda 2800$, H$\gamma$ and H$\beta$.  In addition,
narrow, forbidden transitions of oxygen and highly ionized neon are
also evident.  These features, seen in both components of the PA $=60^{\circ}$ 
spectra, clearly confirm the lensing nature of \GRALGL.  There is 
no clear evidence of the lensing galaxy in the Keck data, neither
in the sky-subtracted two-dimensional spectra, nor in the extracted,
calibrated one-dimensional spectra.

%%%%% MODELING %%%%%
\section{Lens modeling}
\label{sec:modeling}

%\noindent
%\textcolor{red}{-- Modélisation avec différents modèles simples (SISg, NIEg, Hern+gNFW+shear) en utilisant (i) l'astrométrie uniquement, (ii) l'astrométrie et la photométrie}

%\noindent
%\textcolor{red}{-- Estimation des délais temporels en fonction du redshift inconnu du déflecteur (courbe de $Delta_t$ en fct de $z_s$), et ce pour chaque modèle}

%\noindent
%\textcolor{red}{-- Estimation de l'impact d'un modèle SPT-modifié sur les délais temporels (cfr Wertz, Orthen \& Schneider, 2018)}

%\noindent
%\textcolor{red}{PLACE HOLDER TEXT A GL  \citep[NSIEg,][]{1994A&A...284..285K,1987ApJ...312...22K}.  massive early-type galaxies in the region where multiple images occur \citep{2017arXiv171204945G}.}

% IDEAS
% What are the observables we use as constraints to fit the model? --> The constraints include the positions of the 4 lensed images and the 3 flux ratios in G-band, both provided in the {\it Gaia} DR2 (ref to paper I). To account for unquantified source of error (reddening, dust, ...) we apply a 10% error on the flux ratio

% Which cosmology? --> Planck15 with H0 = h x 100 km s-^1 Mpc^-1

% explore, describe, investigate, probe, infer, cause

%%% MAIN
%\textcolor{green}{[The idea is to compute the predicted time delays.]}
In this section, we describe the method applied to obtain a simple lens model that can adequately reproduce the lensing observables provided by {\it Gaia} DR2. Our motivation is to use this model to predict time delays between pairs of lensed images for a range of plausible lens redshift values.

% autres techniques tirant profit d'image haute resolution

\subsection{Overview}
\label{subsec:modeling:overview}
% Which software?

% Which observables (--> image positions considered as point-like object + flux ratios), add a table + reference to paper I and III
% images and the flux ratios in $G$-band, each being provided with the {\it Gaia} DR2. 

The constraints on the lens mass distribution include the relative angular positions $\bt_i$ of the lensed images with respect to the brighter image, hereafter labelled  image A, and the flux ratios $f_i \equiv F_i / F_{A}$ in the $G$-band between the images $i$ and A. Because the number of constraints is quite limited, we reconstructed the lens mass distribution using only a simple physically motivated and fully parametrized model, which is described in Sect. \ref{subsec:modeling:models}. 
%\textcolor{green}{[OW: add references to other techniques we will use when high-res observations will be obtained, e.g., Birrer, Amara \& Refregier, (2015).]}

Both the astrometric and photometric {\it Gaia} measurements are affected by statistical errors. %, which are also provided with the {\it Gaia} DR2. 
However, the flux uncertainties as given in {\it Gaia} DR2 do not reflect various well-known sources of uncertainty that have to be taken into account in the modeling scenario, the most important of which are 
(i) microlensing effects of one or several of the macrolensed images \citep[see, e.g.,][]{Wambsganss_Q2237_1991, Chae_H1413_2001, Akhunov_Wertz_2017}, 
(ii) small scale structures in the lens galaxy at the image positions \citep[][and references therein]{Mao_Schneider_1998, Metcalf_Madau_2001, Hsueh_substructure_2017}, 
(iii) differential dust-reddening \citep[see, e.g.,][]{GL_dustreddening_2004, Jean_Surdej_2007, GL_dustreddening_2008}, and 
(iv) source variability which propagates into image light curves with lags due to time delays \citep[][]{Treu_TDcosmography_2016}.
To represent these unquantified effects, we thus used conservative $15\%$ Gaussian errors for the image flux ratios. Both the lensing observables and their related uncertainties are reported in Table \ref{tbl:observables}.

We performed the modeling using \texttt{pySPT} \citep[][]{pySPT_2018}, a software package mainly dedicated to the study of the Source Position Transformation (SPT) but which comes with several simple modeling tools, and \texttt{gravlens}, a lensing-dedicated software package developed by C. R. Keeton \citep[][]{Keeton_modeling_2001, Keeton_modeling_2010, Keeton_gravlens}. 

% Measurement errors --> Gaia | other errors --> flux ratios
% Measurement errors --> may be underestimated the uncertainties affecting the flux ratios.

%To take into account of the various sources of error affecting the image flux measurements
% both reported in Table ??
% To account for unquantified source of error (microlensing, reddening, dust, ...) which might affect the image fluxes, we apply a conservative 10% error on the flux ratio

% strongly constrain the flux ratios, then relaxing the errors

\begin{table}
\caption{{\it Gaia}'s lensing observables for \GRALGL.}
\label{tbl:observables}
\footnotesize
\begin{center}
\begin{tabular}{cccc}
\hline
\hline
Image 	& $\Delta \alpha \cos{(\delta)}$ 	& $\Delta \delta$ 	& flux ratios	\\
 		& ["]								&  ["]				& in $G$-band		\\
\hline
A		& $\,0.00000 \pm 0.0021$ & $ 0.0000 \pm 0.0020 $	& $1.00 \pm 0.15$ \\
B		& $\,\,0.3454 \pm 0.0020$ & $-0.3246 \pm 0.0015 $ 	& $0.95 \pm 0.15$ \\
C		& $-1.2825 \pm 0.0021$ & $-0.4246 \pm 0.0015 $ 	& $0.47 \pm 0.15$ \\
D		& $-0.3434 \pm 0.0023$ & $-1.5110 \pm 0.0015 $ 	& $0.40 \pm 0.15$ \\
\hline
\end{tabular}
\end{center}
\end{table}

\subsection{Lens mass models}
\label{subsec:modeling:models}

% We consider only a single lens galaxy in our model scenario --> the DK1.54m image shows no evidence of other galaxies in the vicinity of \GRALAL

We modeled the mass distribution of the lens galaxy using %two different classes of parametrized density profiles. First, we considered 
a singular power-law elliptical mass distribution (SPEMD), which is %known to provide a proper characterization 
broadly consistent with typical lens galaxies and has been extensively used in the literature \citep[see, e.g.,][]{Suyu_2009, Sonnenfeld_2013, Birrer_RXJ1131_2016, Wong_Holicow_2017, Shajib_2018}. The corresponding dimensionless surface mass density profile, also known as the convergence, is defined by
\begin{equation}
\kappa(\theta_1,\theta_2) = \frac{a}{2} \left(\frac{\tE}{\sqrt{\underline{\theta}_1^2 / q + q\underline{\theta}_2^2}}\right)^{2 - a} \,,
\label{eq:kappa_spemd}
\end{equation}
where $\tE$ is the Einstein radius, $a$ the power-law slope\footnote{The power-law slope $a$ is linked to the 3-dimensional slope $\gamma'$ of the power-law mass distribution through the relation $a = 3 - \gamma'$.}, and $q$ the minor-to-major axis ratio of the elliptical iso-density contours. The on-sky Cartesian angular coordinates $(\theta_1,\theta_2)$ are clockwise rotationally transformed into the coordinates $(\underline{\theta}_1,\underline{\theta}_2)$, whose axes are aligned with the minor and major axes of the lens. Specifically, we write $\underline{\bt} = R(-\tq)\,\bt$ where $R$ is the rotation matrix and $\tq$ the position angle of the minor axis. 
%\begin{equation}
%\begin{pmatrix} \underline{\theta}_1 \\ \underline{\theta}_2 \end{pmatrix} = \begin{pmatrix} \cos{\tq} & \sin{\tq} \\ -\sin{\tq} & \cos{\tq} \end{pmatrix} \begin{pmatrix} \theta_1 \\ \theta_2 \end{pmatrix} \,,
%\label{eq:rotation}
%\end{equation}
%where $\tq$ is the minor axis position angle.
The SPEMD simplifies into a singular isothermal ellipsoid (SIE) when $a = 1$, and into a singular isothermal sphere (SIS) when both $a = 1$ and $q = 1$.
Closed-form expressions for the SPEMD deflection angle $\ba$ and deflection potential $\psi$ can be found in \citet[][]{Keeton_masscatalog_2001}.

%Secondly, we considered a composite model consisting of a Hernquist profile \citep[][]{Hernquist_1990} and a generalized Navarro-Frenk-White profile \citep[][]{NFW_1996} to respectively trace the baryonic matter and the dark matter halo \citep[see, e.g.,][and Appendix \ref{appendix:composite} for more details]{Courbin_2011, Schneider_Sluse_2013, Xu_2013}.

%In both cases, 
We also included external shear \citep[see, e.g.,][]{Saas_GL_2006} to describe the weak influence of long-scale structure and possible local massive objects. This adds the two parameters $\gamma$ (shear strength) and $\tg$ (shear position angle) to the lens model. In addition to the model parameters, both the position of the source $(\beta_1,\beta_2)$ and the lens galaxy centroid $(\xgal,\ygal)$ are also unknown quantities. A first estimate may however be inferred from the centroid of the four lensed image positions, namely $(c_1,c_2) = (-0.\!\!^{\prime\prime}320,-0.\!\!^{\prime\prime}565)$ with respect to the image A, which has been used as a prior for both $(\bx,\by)$ and $(\xgal,\ygal)$.

\subsection{Modeling procedure}
\label{subsec:modeling:procedure}

For a given lens model, the parameter space is in general populated with several local solutions which optimize the objective function in some feasible neighborhood. As a first step, we explored the parameter space using the differential evolution algorithm \citep[][]{Storn_DE_1997}, which is designed to search for the global solution with no absolute guarantee to find it. As a benefit, this method requires no initial solution, but ranges of parameter values only. %, we provide the parameter ranges we used for each model. 
We defined half the largest angular separation $\tmax$ between the images as a prior for $\tE$, and considered the initial range $[\tmax/4, 4\tmax]$. For both the source position and lens galaxy centroid, we set the initial range $[c_j-\tmax/2, c_j+\tmax/2]$ for each coordinate ($j = 1, 2$). We also set the initial ranges $q \in [0.1,1.0]$, $\gamma \in [0.0, 0.3]$, $a \in [0.5, 1.5]$, and $[0, 2\pi]$ for the two angular parameters $\tq$ and $\tg$.

For this first step, we ignored the observed flux ratios and only minimized the dispersion of the sources $\bb_i \equiv \bb(\subb{\bt}{obs}{i},\bp) = \subb{\bt}{obs}{i} - \ba(\subb{\bt}{obs}{i})$, which are traced back from the observed image positions $\sub{\bt}{obs}$ for a given set of parameters $\bp$. Because it does not require the lens equation to be solved, calculating $\bb_i$ is very efficient. We obtained a first set of parameters $\bp_0$, which includes the SPEMD model parameters, the lens galaxy centroid, and the source position $\bb_0$ derived from the median value of $\bb_i$ resulting from the best fit. To decrease the chance of getting stuck in a local solution, we also ran the minimization process on sub-regions of the parameter space, and compared the sub-results with the one obtained when running on the entire parameter space. We thus used the flux ratios as an additional constraint to separate plausible from poor solutions. 

As a second step, we refined the solution using a downhill simplex algorithm \citep[][]{NelderMead65} with $\bp_0$ as the first guess. Assuming that the errors follow a Gaussian distribution, %Gaussian data errors, 
the goodness of the fit was evaluated with a reduced $\chi^2$ statistics, $\sub{\chi}{red}^2 = \chi^2 / \sub{N}{dof}$, where $\sub{N}{dof}$ represents the number of degrees of freedom, naively defined as the difference between the number of lensing observables and that of the model free parameters \citep[][]{Andrae_chi2_2010}. The $\chi^2$ statistic results from the sum of the two contributions $\sub{\chi}{img}^2$ and $\sub{\chi}{flux}^2$, characterized by
\begin{equation}
\chi^2 =\sum_{j=1}^{\sub{N}{img}-1} \frac{\left(\subb{\theta}{obs}{j} - \theta_j\right)^2}{\sigma_{\theta,j}^2} + \sum_{j=1}^{\sub{N}{img}-1} \frac{\left(\subb{f}{obs}{j} - f_j\right)^2}{\sigma_{f,j}^2}\,,
\label{eq:chi2img}
\end{equation}
where $\sub{\bt}{obs}$ and $\bt$ are respectively the observed and model-predicted positions of the lensed images, $\sub{f}{obs}$ and $f$ are the observed and model-predicted flux ratios, and $\sigma$ their associated uncertainties. 
%, and $\sub{N}{dof}$ the number of degrees of freedom. 
%The latter is naively defined as the difference between the number of lensing observables and that of the model free parameters \citep[][]{Andrae_chi2_2010}.
%Although more reliable methods rather than a $\chi^2$ statistic exist \citep[see, e.g.,][]{Andrae_chi2_2010}, such a choice will be sufficient for our purposes.
%resulting 

% chi2 decomposed into img + flux
%we used $(\bp_0, \bb_0)$ as the initial solution 
% we run the algo on subdivided regions of the parameter space
%we also divided the parameter space in numerous regions 
%The range of angular parameters were defined to $[0,\pi]$, the shear strength to $[0.0, 0.2]$, and the lens galaxy centroid components to $[c_x - 0.\!\!^{\prime\prime}65, c_x + 0.\!\!^{\prime\prime}65]$ and $[c_y - 0.\!\!^{\prime\prime}65, c_y + 0.\!\!^{\prime\prime}65]$.
%the Einstein radius to $[0.\!\!^{\prime\prime}3, 1.\!\!^{\prime\prime}5]$
% physically meaningful
% what about the ranges?
% for critical parameters which are known to be easily degenerated
% diff_evo and restricting the range for some parameters
% To decrease the chance of getting stuck on a poor local solution (premature convergence)

%For both the SPEMD and composite models, 
%some parameters (e.g., SPEMD with $a=0$ and $q=1$) 
We initiated the modeling procedure by fixing the parameters $a=0$ and $q=1$, and successively increased the model complexity. We tested the statistical significance of including these parameters using an $F$-test \citep[see, e.g.,][]{Bevington_1969}. %, which helps to decide which model describes the data in a better way. 
Following \citet[][]{Cohn_2001} and \citet[][]{Protassov_Ftest_2002}, we recall that adding a parameter to a given model is statistically significant, with a confidence level $0<\nu<1$ compared to the improvement expected for a random variable, if the difference between the $\chi^2$ statistics, $|\Delta \chi^2| > h\ \chi^2 / \sub{N}{dof}$, exceeds the reduced $\chi^2$ obtained for the unmodified model by a factor $h = \text{ppf}(\nu, \Delta\sub{N}{dof}, \sub{N}{dof}) / \Delta\sub{N}{dof}$, where ppf is the so-called percent point function of the $F-$distribution \citep[][]{David_Fdistribution_1949, Pearson_1950}. For instance, if adding one parameter to a model having $\sub{N}{dof} = 4$ improves the fit from $\chi^2=21.1$ to $3.5$, one obtains $|\Delta \chi^2| = 17.6 \geq 6.79 = (21.1/4) \times \text{ppf}(0.68, 1, 4)$, and the $F-$test suggests that adopting the new model is statistically justified under the $1\sigma$ confidence level hypothesis. Similarly, one can compute the confidence level $\nu^{\star}$ for which $|\Delta \chi^2| = h\ \chi^2 / \sub{N}{dof}$, and compare it with the $1\sigma$ confidence level $\nu_{1\sigma} \simeq 0.68 = 68\%$. In our example, one has $\nu^{\star} \simeq 86\% > \nu_{1\sigma}$, which leads to the same conclusion.

% On peut aussi déterminer nu tel qu'on ait l'égalité, puis comparer avec 0.68 (1sigma confidence level)

%the new model 

%statistically significant if $a>b$ at a $1\sigma$ confidence level.

% is statistically justified

% which is < than the 1sigma confidence level $\nu_{1\sigma} = 0.68$

%we improve a fit by $|\Delta \chi^2| = 0.5$ by adding $1$ parameter to model characterized with $\sub{N}{dof} = 4$

%for a given $\chi^2$ improvement, say $|\Delta \chi^2| = 0.5$

% $1\sigma$ statistically significant

% adding a random variable
% and  defines 
%for a significance level $0<\nu<1$ compared to the improvement expected for a random variable.
% In other words, adding a parameter is statistically significant if ++(>)-formule with h++ for a ??% significance level compared to the improvement expected for a random variable,
% , with a significance level of $0 < \varepsilon < 1$,
% \varpesilon the significance levels compared to the improvement expected for a random variable
%$|\Delta \chi^2| > \text{ppf}(\varepsilon, \Delta\sub{N}{dof}, \sub{N}{dof})\ \chi^2 / \sub{N}{dof}$, where ... is the percent point function for the F-distribution

%For the model describing the data in a better way from a statistical point of view
As a third step, we further explored the parameter space of the best fit model and deduced confidence intervals for each model parameter using a Bayesian inference method based on a Monte-Carlo Markov Chain (MCMC) sampling. We sampled the posterior probability distribution function (pdf) of the model parameters using \texttt{emcee} \citep[][]{emcee}, a \texttt{Python} package which implements the affine-invariant ensemble sampler for MCMC proposed by \citet[][]{Goodman_MCMC_2010}. We monitored the chains, also called walkers, using the fraction of accepted to proposed candidates \citep[the so-called acceptance rate,][]{Mackay_2003}, and assessed the convergence with the integrated autocorrelation time \citep[][]{Christen_Fox_2010,Goodman_MCMC_2010}, which gives an estimate of the number of posterior pdf evaluations required to draw an independent sample. We then computed the $1\sigma$ confidence intervals from the $16$- and $84$-th percentiles.

%The convergence has been typically reached after a few thousands iterations. 

%To carry out this analysis, we used the \texttt{emcee} package \citep[][]{emcee} written in \texttt{Python}, 

% blue regions --> combined errors

\subsection{Model properties}
\label{subsec:modeling:properties}

% Different single solutions for each model + statistical analysis (F-test)
% Adopt the SIEg for the MCMC

\begin{table*}
\caption{The best-fit model parameters obtained for the three complexity levels of the SPEMD plus external shear. The confidence intervals of the SIEg parameters, inferred from the MCMC sampling, are also reported at the bottom.}
\label{tbl:ranges}
\footnotesize
\begin{center}
\begin{tabular}{lccccccccccc}
\hline
\hline
% add the chi2 results
Model 	& $\sub{\chi}{red}^2$ & $\tE$ 	& $a$ 	& $q$ & $\tq$ & $\gamma$ & $\tg$ & $\bx$ & $\by$ &$\xgal$ & $\ygal$	\\
% 		& ["]								&  ["]				& in $G$-band		\\
\hline
SISg   & $90.6/4$ & $\arc{0.}{851}$ & $\equiv 1.0$ & $\equiv 1.0$ & $-$               & $0.048$ & $\dek{-3.}{7}$ & $\arc{-0.}{445}$ & $\arc{-0.}{705}$ & $\arc{-0.}{427}$ & $\arc{-0.}{737}$\\
SIEg   & $ 1.2/2$ & $\arc{0.}{851}$ & $\equiv 1.0$ & $0.915$      & $\dek{150.}{7}$&  $0.044$ & $\dek{11.}{5}$  & $\arc{-0.}{447}$ & $\arc{-0.}{704}$ & $\arc{-0.}{424}$ & $\arc{-0.}{744}$\\
SPEMDg & $ 1.1/1$ & $\arc{0.}{853}$ & $1.299$      & $0.885$      & $\dek{2.}{0}$& $0.023$ & $\dek{0.}{6}$  & $\arc{-0.}{440}$ & $\arc{-0.}{719}$ & $\arc{-0.}{422}$ & $\arc{-0.}{752}$\\
\hline
C.I. & &$\arc{0.}{851}_{-0.001}^{+0.002}$& $\equiv 1.0$ & $0.914_{-0.008}^{+0.007}$ & $\dek{151.}{4}_{-3.0}^{+2.5}$& $0.044_{-0.002}^{+0.002}$ & $\dek{11.}{7}_{-1.1}^{+1.4}$ & $\arc{-0.}{447}$ & $\arc{-0.}{704}$ & $\arc{-0.}{426}$ & $\arc{-0.}{743}$ \\
\hline
\end{tabular}
\end{center}
\end{table*}

We first examined the SPEMD with $a=1$ and $q=1$ in an external shear field, hereafter denoted as the SISg model. This model is characterized by seven parameters $(\tE, \gamma, \theta_\gamma, \bx, \by, \xgal, \ygal)$ and $\Ndof = 4$. With $\chi^2 / \Ndof = (89.3 + 1.3) / 4$, the SISg model poorly fits the data, in particular the image positions. Although this result comes as no surprise considering the incredible simplicity of the SISg model, it provides a first estimate of the Einstein radius, $\tE = \arc{0.}{851}$, and of the external shear, $(\gamma, \theta_\gamma) = (0.048,\dek{-3.}{7})$, see Table \ref{tbl:ranges}. Adding an ellipticity $(q,\tq)$ parameter to the SISg, thus transforming it into a SIEg, considerably improves the fit to $\chi^2 / \Ndof = (0.0 + 1.2) / 2$, and is statistically significant with a confidence level $\nu^{\star} = 75\%\ (> 68\%)$ for the $F-$test. Letting the parameter $a$ vary during the optimization process, we relaxed the isothermal hypothesis. This slightly improves the fit, $\chi^2 / \Ndof = (0.0 + 1.1) / 1$, but is not statistically significant according to the $F$-test ($\nu^{\star} = 28\%$). Increasing the model complexity favors slightly higher ellipticity $(1-q)$ along with smaller external shear strength. This clearly reflects the well-known degeneracy existing between these two sources of angular structure \citep[see, e.g.,][]{Keeton_shearellipticity_1997, Keeton_modeling_2010, Kneib_Natarajan_2012}.

\begin{figure}[!htp]
\begin{center}
\includegraphics[width=0.48\textwidth]{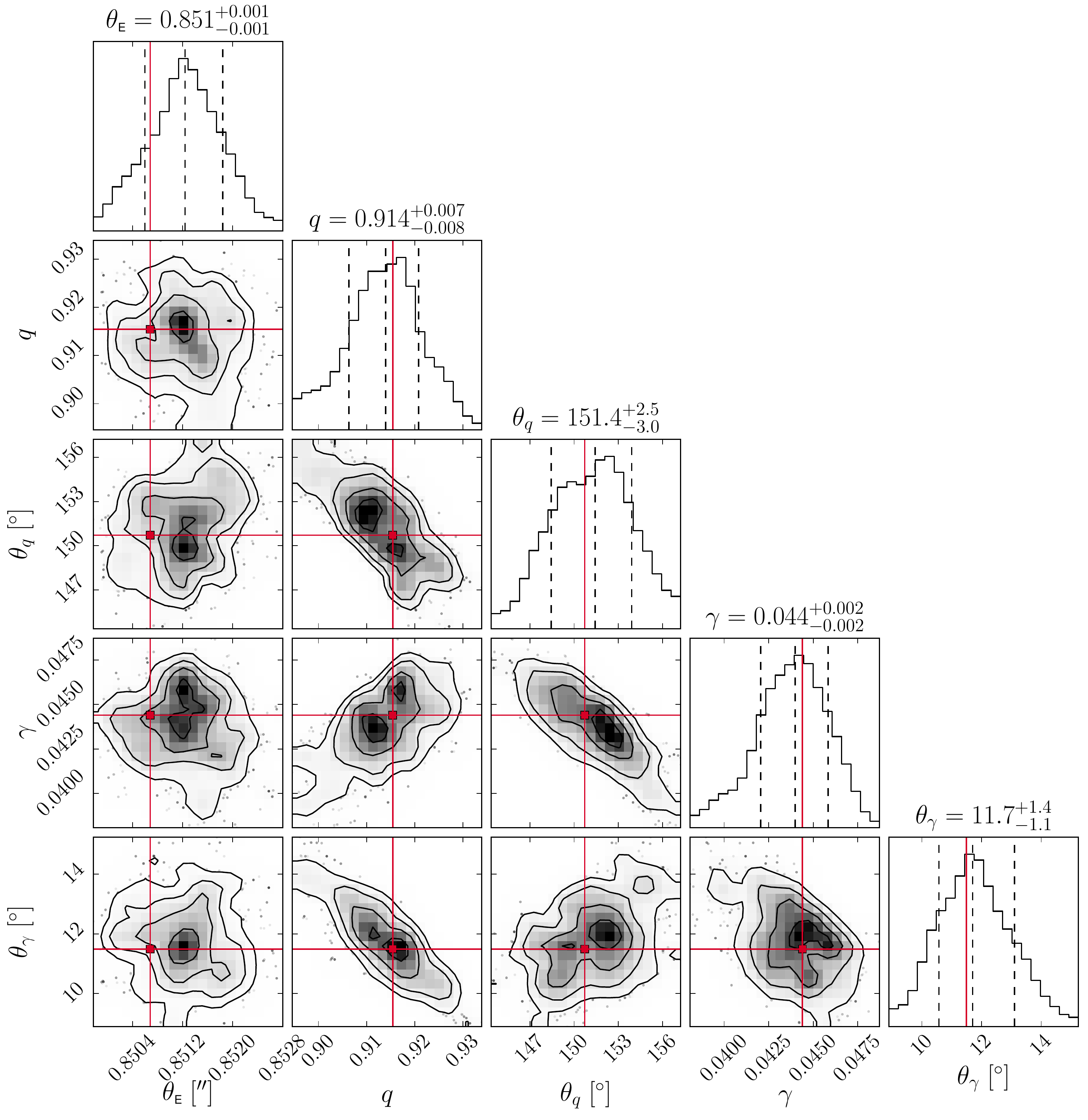}
\caption{Results of MCMC sampling for the SIEg model parameters. The diagonal panels illustrate the posterior pdfs while the off-axis ones illustrate the correlation between the parameters. The vertical red lines and red crosses correspond to the best solution obtained from the down-hill simplex algorithm and used to initiate the $250$ walkers. The vertical dashed lines locate the $16-$ and $84-$th percentiles. 
\label{fig:mcmc1}}
\end{center}
\end{figure}

\begin{figure}[!htp]
\begin{center}
\includegraphics[width=0.48\textwidth]{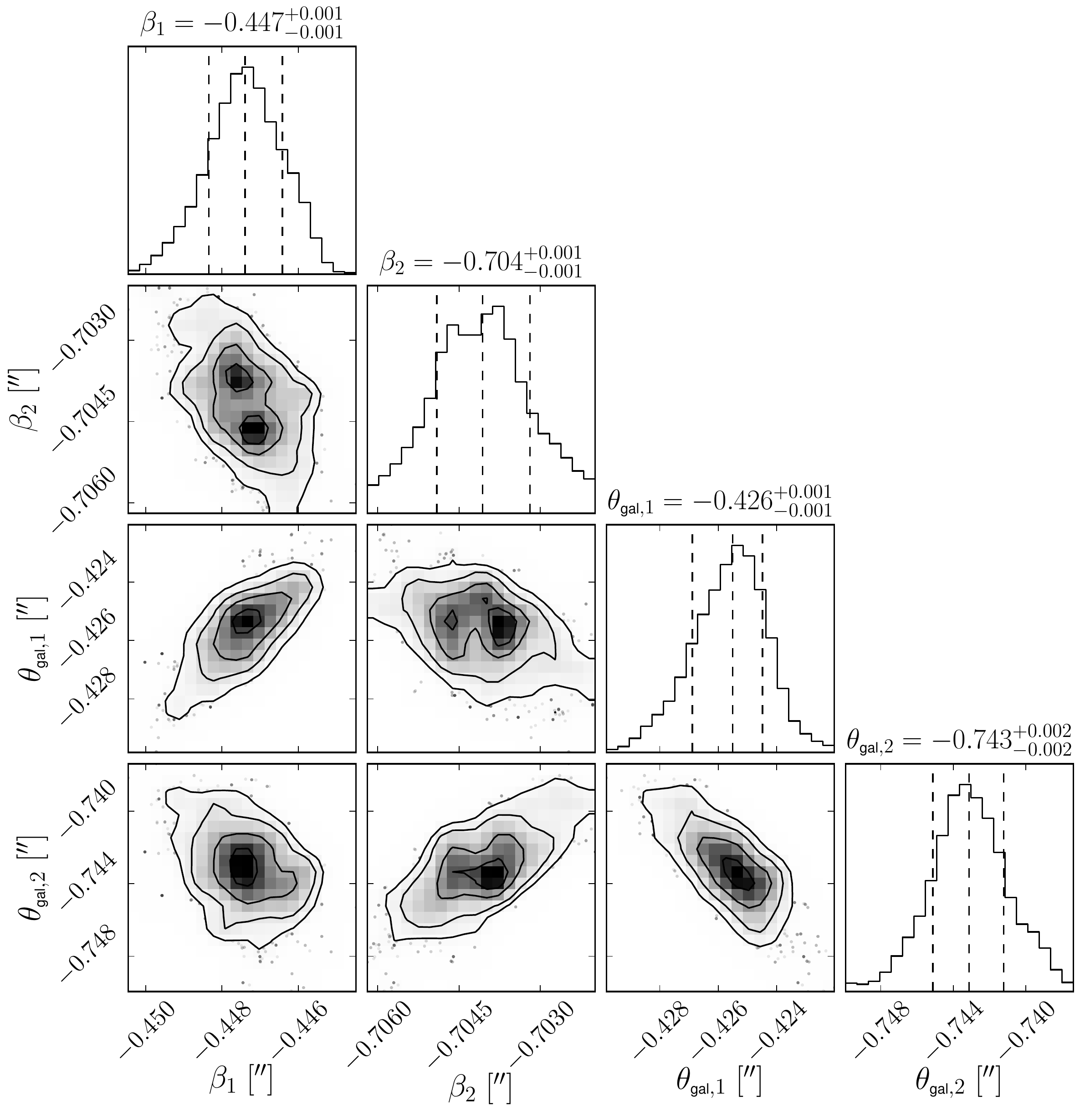}
\caption{Results of the MCMC sampling for the source position and lens galaxy centroid. The diagonal panels illustrate the posterior pdfs while the off-axis ones illustrate the correlation between the parameters. The vertical dashed lines locate the $16-$ and $84-$th percentiles.
\label{fig:mcmc2}}
\end{center}
\end{figure}

\begin{figure}[!htp]
\begin{center}
\includegraphics[width=0.48\textwidth]{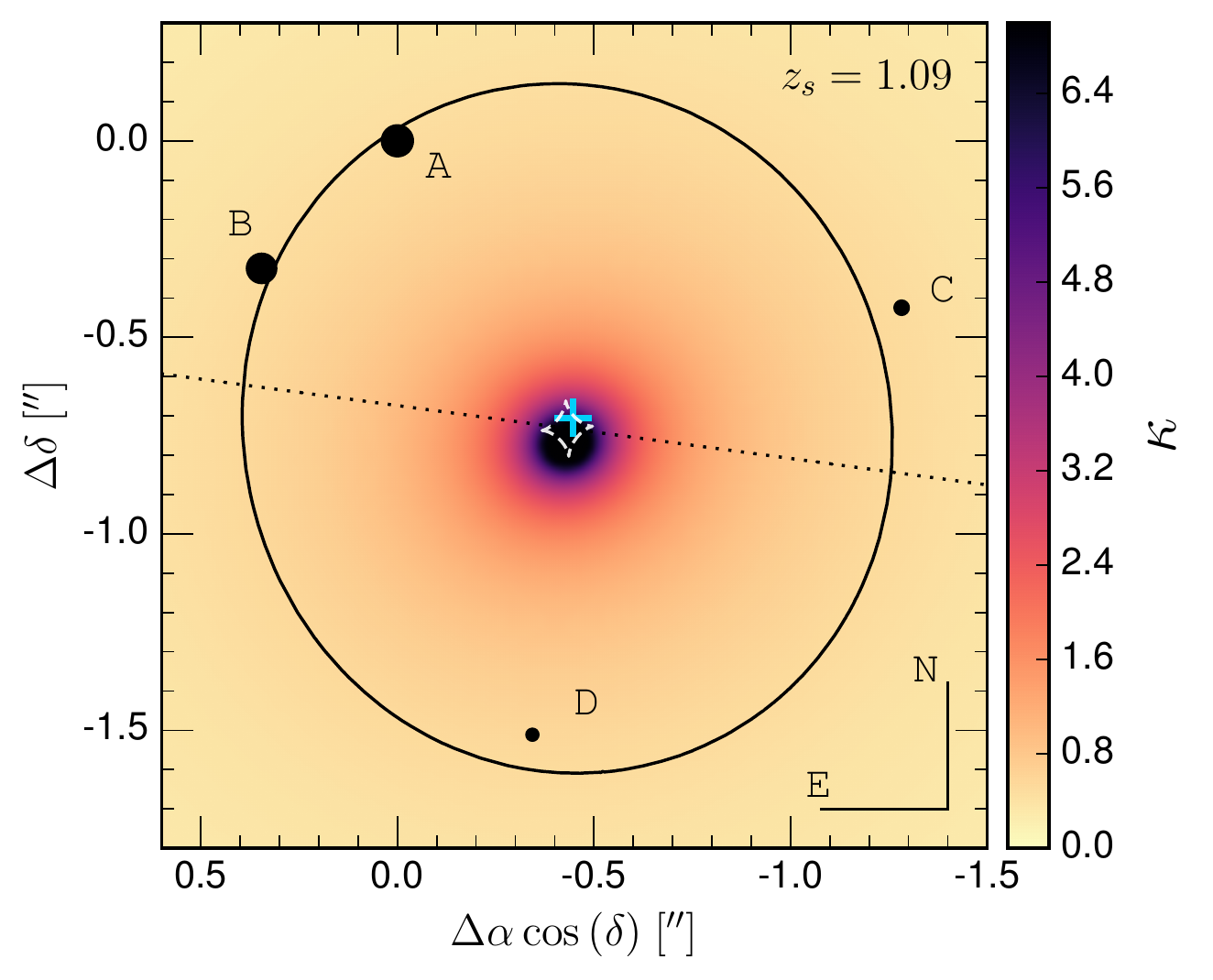}
\caption{The GRAL113100-441959 image configuration. The black dots locate the image positions, and their size mimics the associated flux, as reported in the {\it Gaia} DR2. The solid line represents the tangential critical line, the diamond-shaped dashed line represents the corresponding caustic line, and the dotted line defines the direction of the external shear. Finally, the color map shows how the surface mass density $\kappa$ is distributed.
\label{fig:lensplane}}
\end{center}
\end{figure}

The errors $\sigma_{f}$ adopted for the image flux ratios are to some extent arbitrary. We thus explored the impact on the fit when we modify these values. In this regard, we restrained ourselves to estimate these errors by rescaling $\sigma_{f}$ to artificially obtain $\sub{\chi}{red}^2 = 1$, because this method has been shown to be incorrect \citep[e.g.,][]{Andrae_errorestimation_2010}. We found that the SIEg remains the one providing the most statistically significant results. As a next step, we sampled the pdfs for all SIEg parameters, using the best fit parameter values reported in Table \ref{tbl:ranges} to initialize $250$ walkers. The resulting pdfs and the correlation between model parameters are represented with two corner plots, in Figs. \ref{fig:mcmc1} and \ref{fig:mcmc2} respectively. The corresponding confidence intervals are also reported in Table \ref{tbl:ranges}.
As expected, the shear-ellipticity degeneracy induces a significant correlation between the parameters $(\gamma, \theta_\gamma)$ and $(q, \theta_q)$. 
In Fig. \ref{fig:lensplane}, we represent the lensed image configuration, labelled from A to D, on top of a few predicted lensing quantities for the SIEg model reported in Table \ref{tbl:ranges}. The predicted mass within a circular aperture of radius $\tE$ is estimated to be in the range $M(\leq \tE) = 2.498 \pm 0.003 \times 10^{11} M_{\odot}$ for $\zl = 0.5$ and $M(\leq \tE) = 1.119 \pm 0.002 \times 10^{12} M_{\odot}$ for $\zl = 0.9$. 

%\footnote{A walker can be roughtly compared to a Metropolis-Hastings chain.}.

% [which err_fluxes to obtain chi2 = 1? Anyway, rescaling the err_fluxes such that chi2_red equals 1 should not be used to infer the err_fluxes --> Andrae (2010) already showed that this method is incorrect \citep[][]{Andrae_errorestimation_2010}]

%Furthermore, we also added a core radius to .... by adding a core radius

% Letting the parameter $a$ varying during the fit only improve the fit by a $\Delta \chi^2 \simeq 0.1$, while adding 1 dof --> does not significantly improve the fit (see "Constraints on Galaxy Density Profiles from Strong Gravitational Lensing: The Case of B 1933+503", Cohn, Kochanek, McLeod, Keeton.
% [use the same argument for SIS --> SIE, but now it makes sense to add q and theta_q]

%%%%% TIME DELAYS %%%%%
\section{Model-predicted time delays}
\label{sec:tds}

%"It is worth noting that time delay perturbations, due to small scale structure in lens plane or along line of sight, will likely not be a significant additional source of systematic error, since they primarily cause additional scatter on hour-long timescales" -- Keeton and Moustakas, 2009.

\begin{figure}[!htp]
\begin{center}
\includegraphics[width=0.48\textwidth]{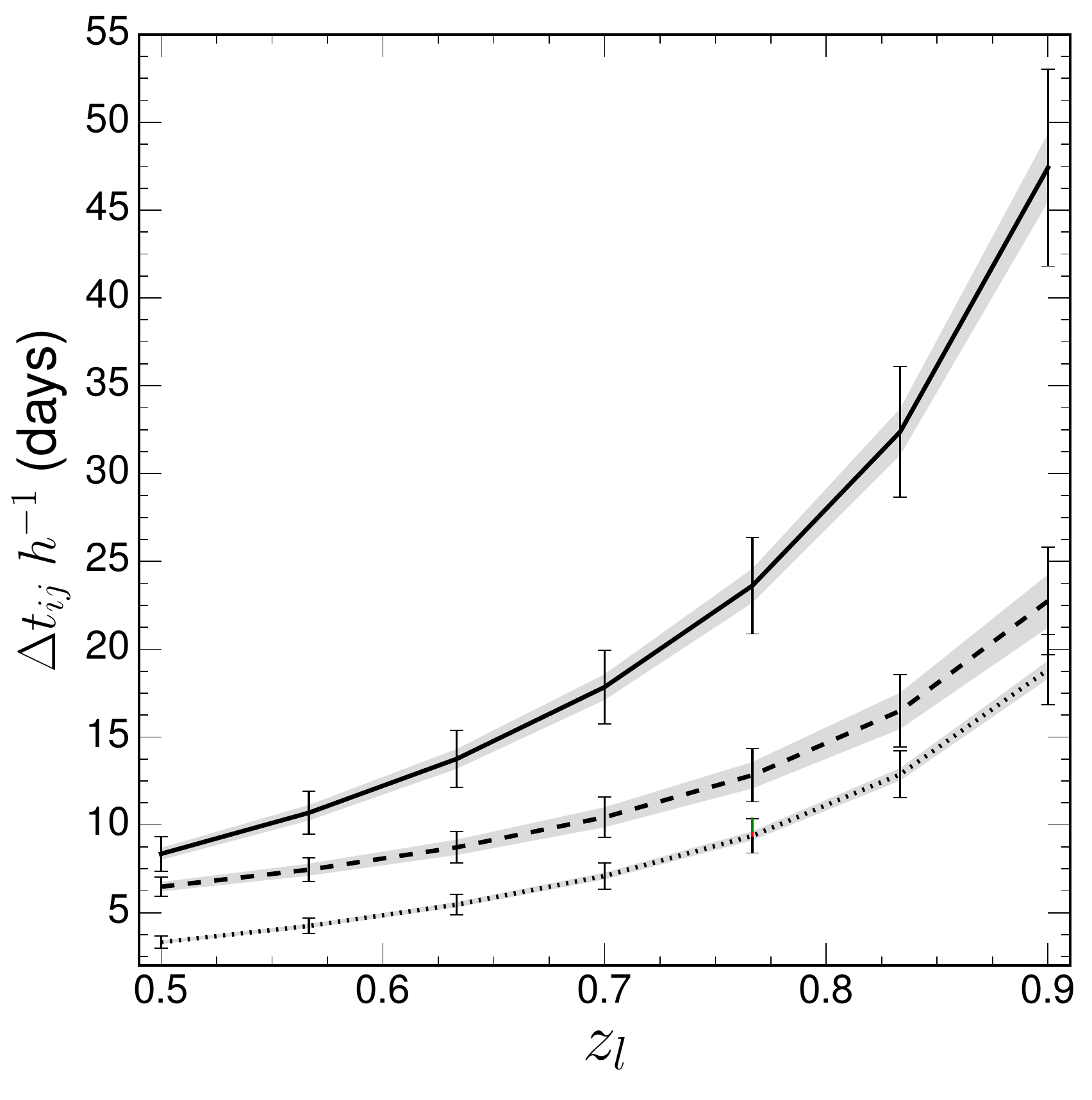}
\caption{Model-predicted time delays scaled with $h$. The error bars combine the three sources of uncertainties described in the text. The grayshaded regions depict the contribution of the SPT to the error budget. The solid line corresponds to images C$-$D, the dashed line to C$-$A, and the dotted line to C$-$B. For the sake of clarity, we shifted upwards the time delays between images C$-$A (dashed line) by a $3$ day offset.
\label{fig:predicted_time_delays}}
\end{center}
\end{figure}

From the set of highly probable lens models obtained from the MCMC sampling, we computed the predicted time delays $\Delta t_{ij}$ between pairs of images $(\bt_i, \bt_j)$, which are defined by
\begin{equation}
\Delta t_{ij} = \frac{D_{\Delta t}}{c}\left[\frac{1}{2}\left(|\ba(\bt_i)|^2-|\ba(\bt_j)|^2\right) - (\psi(\bt_i) - \psi(\bt_j)) \right] \ ,
\label{eq:TD}
\end{equation}
where $D_{\Delta t} = (1+z_l) D_l D_s / D_{ls} \propto H_0^{-1}$ is the time-delay distance, with $D$ the angular diameter distance between the observer and lens ($D_l$), observer and source ($D_s$), and lens and source ($D_{ls}$). The angular diameter distance depends only on the redshift and the cosmology. %, in particular it depends linearly on $H_0^{-1}$. 
From the Keck/LRIS spectra, the redshift of the source was found to be $\zs = 1.09$. However, as there is no clear evidence of the lensing galaxy in the Keck/LRIS spectral data, we were prevented from determining its redshift $\zl$ (see Sect. \ref{subsec:lensing_nature}). We thus computed the time-delay distance, hence the time delays, for a set of lens redshifts in the plausible range $\zl \in [0.5,0.9]$. We adopted the $\Lambda$CDM model along with the final Planck 2018 results \citep[][]{Planck2018}. The normalized $\Delta t_{ij}\ h^{-1}$ against the lens redshift $\zl$ is shown in Fig.\ref{fig:predicted_time_delays}, where the factor $h$ is used to calibrate the Hubble constant $H_0 = 100\ h\ \text{km}\ \text{s}^{-1}\ \text{Mpc}^{-1}$.

\begin{figure}[!htp]
\begin{center}
\includegraphics[width=0.48\textwidth]{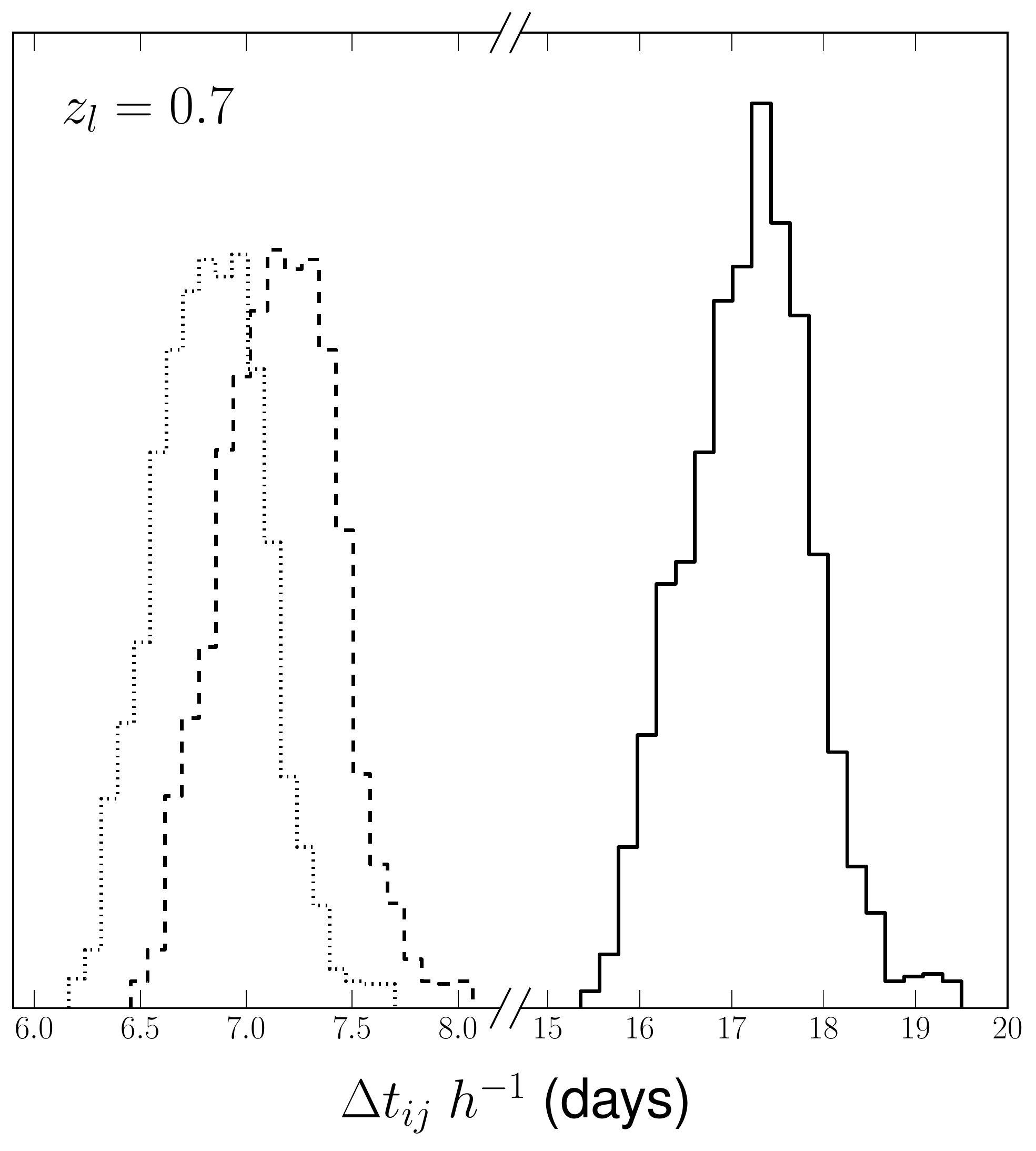}
\caption{Example of histograms for $\Delta t_{ij}\ h^{-1}$ constructed for $\zl = 0.7$ from the MCMC results. The solid line corresponds to images C$-$D, the dashed line to C$-$A, and the dotted line to C$-$B.
%Give the sigma and the corresponding % wrt the median
\label{fig:td_histograms}}
\end{center}
\end{figure}

The error budget associated with $\Delta t_{ij}$ was estimated by considering three sources of uncertainties. First, we propagated the statistical error inferred from the MCMC sampling. This was simply done by constructing histograms for $\Delta t_{ij}\ h^{-1}$ from the independent samples of SIEg model parameters reported in Figs.\ref{fig:mcmc1} and \ref{fig:mcmc2}. An example of these histograms is illustrated in Fig.\ref{fig:td_histograms} for the case of $\zl = 0.7$.
Secondly, we considered the impact of hypothetical massive objects lying on the line-of-sight by scaling the theoretical time-delay distance with an external convergence term $\kappa_{\text{ext}}$ such that $D_{\Delta t} = D_{\Delta t}^{\text{theory}} / (1 - \kappa_{\text{ext}})$ \citep[see, e.g.,][]{Keeton_2003_kappaext}. We applied a different scaling for each set of model parameters used to construct the histograms (see Fig.\ref{fig:td_histograms}). The $\kappa_{\text{ext}}$ values were randomly drawn from a zero mean normal distribution and characterized by a conservative standard deviation $\sigma_{\kappa} = 0.03$ \citep[see, e.g.,][]{Wong_Holicow_2017}. 
When combined, the typical errors are $\sim7.7\%$ for $\Delta t_{\text{CB}}$, $\sim7.8\%$ for $\Delta t_{\text{CA}}$, and $\sim7.6\%$ for $\Delta t_{\text{CD}}$, compared to the median values.
Thirdly, we considered the impact of the Source Position Transformation (SPT), a degeneracy existing between different lens density profiles that reproduce equally well the lensing observables, except for the product $\Delta t\ H_0$ \citep[][]{SPT_SS13, SPT_SS14}. A short summary is given in Appendix \ref{app:spt}. We SPT-transformed the SIEg model using the modified deflection law $\bha = \ba_{\text{SIEg}} + \bb - \bhb(\bb)$ along with a radial stretching of the source plane defined by $\bhb(\bb) = [1 + f_2 |\bb|^2 / (2 \tE)]\ \bb$, where $f_2$ is named the deformation factor. As defined, $\bha$ is not a curl-free field, and hence does not correspond to the deflection produced by a gravitational lens. 

To overcome this hurdle, one can (i) derive the closest curl-free approximation to $\bha$ in a circular region of the lens plane as proposed in \citet[][]{SPT_USS17} and applied in \citet{SPT_WOS_2017} and \citet{pySPT_2018}, or (ii) extract the curl-free part from $\bha$ using an Helmholtz-Hodge decomposition \citep{Helmholtz1858}. In both cases, the new deflection law is denoted $\bta$. We adopted the Helmholtz-Hodge decomposition strategy, given that the region of interest can be straightforwardly reduced to the annulus that includes all four images, leading to higher predicted time delay deviations than found with the first approach (Wertz \& Schneider, in prep.). The higher the value of $f_2$, the larger deformation of the source plane along with the deflection law $\bta$, hence the larger deviations of the predicted image positions in comparison with the observations. For each set of the SIEg model parameters, we thus computed the highest acceptable $f_2$ value, which guarantees the corresponding SPT-transformed model to produce an image configuration identical to the one observed by {\it Gaia}, within the astrometric and photometric error bars. For the best fit SIEg model reported in Table \ref{tbl:ranges}, we found $|f_2|=1.67$, which is representative of the values we obtained for the entire sample. We then repeated the process for the different lens redshifts. Finally, the impact of the SPT results in additional time delay deviations of $2.7\%$ for $\Delta t_{\text{CB}}$, $7.7\%$ for $\Delta t_{\text{CA}}$, and $4.1\%$ for $\Delta t_{\text{CD}}$. This contribution corresponds to a significant fraction of the time delay error budget, in particular for $\Delta t_{\text{CA}}$ where the SPT input reaches the same level as the statistical errors. We finally combined the three different sources of uncertainties to obtain the error bars of Fig.\ref{fig:predicted_time_delays}. 

As a result, the model-predicted time delays vary from (for $\zl = 0.5$)
$\Delta t_{\text{CA}} h^{-1} = 3.48 \pm 0.54$ days, 
$\Delta t_{\text{CB}} h^{-1} = 3.32 \pm 0.35$ days, and 
$\Delta t_{\text{CD}} h^{-1} = 8.35 \pm 0.98$ days, to ($\zl = 0.9$)
$\Delta t_{\text{CA}} h^{-1} = 19.75 \pm 3.07$ days, 
$\Delta t_{\text{CB}} h^{-1} = 18.83 \pm 2.00$ days, and 
$\Delta t_{\text{CD}} h^{-1} = 47.42 \pm 5.61$ days.

% as illustrated with the gray filled regions in Fig.\ref{fig:predicted_time_delays}
% region of interest --> region constrained by the presence of lensed images.
%For $\tE = \arc{0.}{851}$, we found $|f_2|=1.67$ to be the highest acceptable value for the deformation factor. We used that value 
% along with slightly different positions for the lensed images.  undistinguishable image configuration (within the astrometric error bars) compared with the observations.
% The time delay offsets induced by the SPT were combined to ... 
% The time delays offsets induced by the SPT can 
% Each set of model parameters defines a model which can be SPT-transformed...
% plus grande DT, mais on perd la globalité.
%f2 has been chosen as the highest value which leads to undistinguishable image configuration
% allowing values up to 0.1 to a 2\sigma level.

% SPT --> freedom in selecting a model --> density profile only constrained in region where lensed images exist.
 
%%%%% CONCLUSION %%%%%
\section{Conclusions}
\label{sec:conclusions}

% Mentionner les GL confirmés dans la conclusion

This paper presents the spectroscopic confirmation of the gravitationally 
lensed quasar \GRALGL, previously identified in paper I as a highly probable GL candidate hidden in the {\emph Gaia} DR2. %, the first ever lens exclusively discovered from an astrometric survey, here with ESA/Gaia. 
We obtained Keck/LRIS spectroscopy on the night of UT 2018 May 13, revealing similar spectra for the combinations of sources A+B and C+D, for the slit at a position angle of $60^{\circ}$, and A+B and C+D for the slit at a position angle of $135^{\circ}$. We confirmed the lensing nature of \GRALGL\ by clearly identifying 
several similar emission spectral lines between the images (A+B) and (C+D), and measured redshits for these combination of lensed images of the quasar as $\zs = 1.09$. 
%of a unique source for the images ? and ? \textcolor{red}{[OW --> DS: Which images?]}.

We modeled the main lens galaxy with a SPEMD plus external shear, and explored different levels of model complexity. This study indicates that the 
SIEg model (which is equivalent to SPEMD with $a=1$) is the best choice in terms of goodness of the fit and statistical significance, given the available observational data. We inferred the confidence intervals for each model parameter using a Bayesian inference method based on a MCMC sampling. We found $\tE = \arc{-0.}{851} \pm 0.001$ for 
the Einstein radius. The lens galaxy ellipticity is inferred to have an axis ratio of $q = 0.914^{+0.007}_{-0.008}$ pointing to $\dek{151.}{4}^{+2.5}_{-3.0}$ 
east of north. The shear strength is found to be moderate, $\gamma = 0.044 \pm 0.002$, pointing $\dek{11.}{7}^{+1.4}_{-1.1}$ east of north. This suggests that the ellipticity captures most of the source of angular structure. Finally, we predict the lensing galaxy to lie at $(\Delta \alpha \cos(\delta), \Delta \delta)=(\arc{-0.}{426} \pm 0.001, \arc{-0.}{743} \pm 0.002)$ from image A.

We also used the MCMC results to predict the three independent time delays between pairs of images. As the redshift $\zl$ of the lensing galaxy is still unknown, 
we provide the model-predicted time delays for a range of physically plausible $\zl$, between $0.5$ and $0.9$. Finally, we assessed the error affecting our estimation, and in particular quantified the impact of the SPT. We found that the contribution of the SPT to the error budget cannot be ignored for precise applications of strongly lensed systems, as it can be as important as the other uncertainty sources. In the case of \GRALGL, SPT is extremely relevant for the time-delay between the C$-$A images.

\GRALGL\ is the first spectroscopically confirmed GL from the Gaia GraL sample of highly probable candidates recently presented in papers I and III of this series. 
At the time we release this work, nine additional candidates from Paper III were already observed with the Keck/LRIS in other campaigns, with six of them being %\textcolor{red}{either} 
very likely %\textcolor{red}{or confirmed} 
lenses; the analysis of these other GL candidates will be presented in a companion paper.

%$6$ out of $9$ candidates observed
%with the Keck/LRIS facility have also been spectroscopically confirmed, and will shortly be
%presented in a companion paper. 
%\textcolor{green}{[OW: Should we explicitely list all the confirmed GL here?]}

% for a total of 10 observed.

% pendant l'écriture de ce papier, 6 autres ont été confirmé, et feront l'objet d'une publication prochaine.

%%%%% ACKNOWLEDGEMENTS %%%%%
% ADD a thanks-comment for Uffe Graae Jorgensen, Martin Dominik and the DK1.54 observers (Martin Burgdorf, ...)
\begin{acknowledgements}
    The authors would like to warmly thank Uffe Gr{\aa}e J{\o}rgensen and Martin Dominik for having freed up observation time with the Danish telescope, and Martin Burgdorf for having observed our target. 
      OW is supported by the Humboldt Research Fellowship for Postdoctoral Researchers.
      The work of DS was carried out at the Jet Propulsion Laboratory,
California Institute of Technology, under a contract with NASA.
       AKM acknowledges the support from the Portuguese Funda\c c\~ao para a Ci\^encia e a Tecnologia (FCT) through grants SFRH/BPD/74697/2010, PTDC/FIS-AST/31546/2017, from the Portuguese Strategic Programme UID/FIS/00099/2013 for CENTRA, from the ESA contract AO/1-7836/14/NL/HB and from the Caltech Division of Physics, Mathematics and Astronomy for hosting a research leave during 2017-2018, when this paper was partially prepared. AKM additionally acknowledges that this research was partially supported by the Munich Institute for Astro- and Particle Physics (MIAPP) of the DFG cluster of excellence ``Origin and Structure of the Universe''.
      LD and JS acknowledge support from the ESA PRODEX Programme `{\it Gaia}-DPAC QSOs' and from the Belgian Federal Science Policy Office.
      SGD and MJG acknowledge a partial support from the NSF grants AST-1413600 and AST-1518308, and the NASA grant 16-ADAP16-0232.
      We acknowledge partial support from `Actions sur projet INSU-PNGRAM', and from the Brazil-France exchange programmes Funda\c c\~ao de Amparo \`a Pesquisa do Estado de S\~ao Paulo (FAPESP) and Coordena\c c\~ao de Aperfei\c coamento de Pessoal de N\'ivel Superior (CAPES) -- Comit\'e Fran\c cais d'\'Evaluation de la Coop\'eration Universitaire et Scientifique avec le Br\'esil (COFECUB).
      This work has made use of the computing facilities of the Laboratory of Astroinformatics (IAG/USP, NAT/Unicsul), whose purchase was made possible by the Brazilian agency FAPESP (grant 2009/54006-4) and the INCT-A, and we thank the entire LAi team, specially Carlos Paladini, Ulisses Manzo Castello, Luis Ricardo Manrique and Alex Carciofi for the support.
      This work has made use of results from the ESA space mission {\it Gaia}, the data from which were processed by the {\it Gaia} Data Processing and Analysis Consortium (DPAC). Funding for the DPAC has been provided by national institutions, in particular the institutions participating in the {\it Gaia} Multilateral Agreement. The {\it Gaia} mission website is:
http://www.cosmos.esa.int/gaia. Some of the authors are members of the {\it Gaia} Data Processing and Analysis Consortium (DPAC).
\end{acknowledgements}

\bibliographystyle{aa}
\bibliography{bibliography} 

%%%%%%%% APPENDIX %%%%%%%%%
\begin{appendix}

%%%%% >> Composite model
%\section{The composite model}
%\label{appendix:composite}
%
%The \citet[][]{Hernquist_1990} 3-dimensional mass profile $\rho_{\text{hern}}$ is defined by 
%\begin{equation}
%\rho_{\text{hern}}(r) = \frac{\rho_{\star}}{(r/r_{\star})(1 + r/r_{\star})^3} %\,,
%\label{eq:hern}
%\end{equation}
%where $\rho_{\star}$ is a characteristic density, $r_{\star}$ a scale radius, and $r^2 = \theta_1^2 + \theta_2^2$. Analytical expressions for the basic lensing quantities can be found in \citet[][]{Keeton_masscatalog_2001}.
%
%The generalized pseudo-NFW 3-dimensional mass profile $\rho_{\text{NFW}}$ \citep[][]{NFW_1996} is a particular case of the cuspy halo model \citep[][]{Munoz_CuspModel_2001} with a logarithm slope $n$, at large radius, equal to $3$, and is expressed as 
%\begin{equation}
%\sub{\rho}{NFW}(r) = \frac{\sub{\rho}{s}}{(r/\sub{r}{s})^{a} [1 + (r/\sub{r}{s})^{2}]^{(3-a)/2}} \,, 
%\label{eq:gNFW}
%\end{equation} 
%where $\sub{\rho}{s}$ is a characteristic density, $\sub{r}{s}$ the scale radius, and $a$ the logarithmic slope of the density profile at small radius. Analytical expressions for the basic lensing quantities can be found in \citet[][]{Keeton_masscatalog_2001} and \citet[][]{pySPT_2018}.

%%%%% >> SPT
\section{The Source Position Transformation}
\label{app:spt}
In this appendix, we summarize the basic principles of the SPT. For a detailed discussion, we refer the reader to \citet[][]{SPT_SS14, SPT_USS17, SPT_WOS_2017, pySPT_2018}.

The relative lensed image positions $\bt_i(\bt_1)$ of a background point-like source located at the unobservable position $\bb$ constitute the lensing observables that we measure with the highest accuracy and precision. When $n$ images are observed, the mapping $\bt_i(\bt_1)$ only provides the constraints 
\begin{equation}
	\bt_i - \ba(\bt_i) = \bt_j - \ba(\bt_j) \ , \qquad \qquad \forall\, 1 \leq i < j \leq n \ ,
	\label{spt_constraints}
\end{equation}
where $\ba(\bt)$ corresponds to the deflection law caused by a foreground surface mass density $\kappa(\bt)$, the so-called lens. 
The SPT addresses the following question: can we define an alternative deflection law, denoted as $\bha(\bt)$, that preserves the mapping $\bt_{i}(\bt_{1})$ for a unique source?  
If such a deflection law exists, the alternative source position $\bhb$ differs in general from $\bb$. Furthermore, it defines the new lens mapping $\bhb = \bt - \bha(\bt)$, which leads to
\begin{equation}
	\bt = \bb + \ba(\bt) = \bhb + \bha(\bt) \ .
	\label{spt_implicit}
\end{equation}
An SPT consists in a global transformation of the source plane formally defined by a
one-to-one mapping $\bhb(\bb)$, unrelated to any physical contribution such as the external convergence. %\citep{SPT_SS13}.
To preserve the mapping $\bt_{i}(\bt_{1})$, the alternative deflection law thus reads 
\begin{equation}
	\bha(\bt) = \ba(\bt) + \bb - \bhb(\bb) = \ba(\bt) + \bb - \bhb(\bt - \ba(\bt)) \ , 
	\label{hat_alpha_definition}
\end{equation}
where in the first step we used Eq.\,\eqref{spt_implicit} and in the last step we inserted the original lens equation. 
As defined, the deflection laws $\ba(\bt)$ and $\bha(\bt)$ yield exactly the same image positions of the source $\bb$ and $\bhb$, respectively. 

Because $\bha$ is in general not a curl-free field, it cannot be expressed as the gradient of a deflection potential caused by a mass distribution $\hkp$.
Provided its curl component is sufficiently small, \citet[][]{SPT_USS17} have established that one can find a curl-free deflection law $\bta$ that 
is similar to $\bha$ in a circular region of the lens plane denoted as $\mathcal{U}$ where multiple images occur. The corresponding similarity criterion reads
\begin{equation}
	 |\Delta \ba(\bt)| \coloneqq |\bta(\bt) - \bha(\bt)| < \sub{\varepsilon}{acc} \ ,
	\label{criterion}
\end{equation}
for $\bt \in \mathcal{U}$. In \cite{SPT_WOS_2017}, the authors hightlight the limitation of this approach and provide two alternative solutions.

\end{appendix}
\end{document}